\documentclass[12 pt,twoside,aps,prd,amsmath,amssymb,
tightenlines,showpacs,showkeys,eqsecnum]{revtex4}
\usepackage{mathptmx}
\usepackage[mathcal]{euscript}
\usepackage{bm}
\usepackage{graphicx}
\usepackage{hyperref}

\renewcommand{\cal}{\mathcal}

\def \myfiguress #1#2#3#4#5#6#7#8
{\begin{figure}[ht]
    \begin{center}
        \includegraphics[width=#2 \textwidth]{#1.eps}
        \hfill
        \includegraphics[width=#6 \textwidth]{#5.eps}
        \parbox[t]{#4\textwidth}{\caption {#3}\label{#1}}
        \hfill
        \parbox[t]{#8\textwidth}{\caption {#7}\label{#5}}
    \end{center}
\end{figure} }

\begin{document}
\title{String cosmological model in the presence of a magnetic flux}
\author{Bijan Saha}
\affiliation{Laboratory of Information Technologies\\ Joint
Institute for Nuclear Research, Dubna\\ 141980 Dubna, Moscow
region, Russia} \email{bijan@jinr.ru}
\homepage{http://www.jinr.ru/~bijan/}

\author{Mihai Visinescu}
\affiliation{Department of Theoretical Physics\\ National
Institute for Physics and Nuclear Engineering\\ Magurele, P. O.
Box MG-6, RO-077125 Bucharest, Romania}
\email{mvisin@theory.nipne.ro}
\homepage{http://www.theory.nipne.ro/~mvisin/}

%\date{\today}}

\begin{abstract}
A Bianchi type I string cosmological model  in the  presence of a
magnetic flux is investigated. A few plausible assumptions
regarding the parametrization of the cosmic string and magneto-fluid
are introduced  and some exact analytical solutions are presented.
\end{abstract}



\maketitle

\bigskip

%%%%%%%%%%%%%%%%%%%%%%%%%%%%%%%%%%%%%%%%%%%%%%%%%%%%%%%%%%%%%%%%%%%%%%%%%%
            \section{Introduction}
%%%%%%%%%%%%%%%%%%%%%%%%%%%%%%%%%%%%%%%%%%%%%%%%%%%%%%%%%%%%%%%%%%%%%%%%%%
Since the observation of the current expansion of the Universe which
has apparently accelerated in the recent past, the anomalies found in
the cosmic microwave background (CMB) and the large structures
observations it becomes obvious that a pure
Friedmann-Lemaitre-Robertson-Walker (FLRW) cosmology should be amended.

Bianchi type I cosmological models are the simplest anisotropic
Universe models playing an important role in understanding essential
features of the Universe. In this class of models it is possible to
accommodate the presence of cosmic strings. In the last time the string
cosmological models have been used in attempts to describe the early
Universe and to investigate anisotropic dark energy component including
a coupling between dark energy and a perfect fluid (dark matter)
\cite{KM}. Cosmic strings are one dimensional topological defects
associated with spontaneous symmetry breaking in gauge theories. Their
presence in the early Universe can be justified in the frame of grand
unified theories (GUT).

The object of this paper is to investigate a Bianchi type I string
cosmological model in the presence of a magnetic flux. The inclusion of
the magnetic field is motivated by the observational cosmology and
astrophysics indicating that many subsystems of the Universe possess
magnetic fields (see e. g. the reviews \cite {GR,W} and references therein).

In the following section we introduce a system of cosmic string and
magnetic field in the Bianchi type I cosmology presenting some of its
general features. In Section III we introduce a few plausible
assumptions usually accepted in the literature and some exact solutions
are produced. In the last section we present some conclusions and
perspectives.

%%%%%%%%%%%%%%%%%%%%%%%%%%%%%%%%%%%%%%%%%%%%%%%%%%%%%%%%%%%%%%%%%%%%%%%%%%%%%
    \section{Fundamental Equations and general solutions}
%%%%%%%%%%%%%%%%%%%%%%%%%%%%%%%%%%%%%%%%%%%%%%%%%%%%%%%%%%%%%%%%%%%%%%%%%%%%
We consider the gravitational filed given by an anisotropic
Bianchi type I (BI) metric
\begin{equation}
ds^2 = a_0^2 (dx^0)^2 - a_1^2 (dx^1)^2 - a_2^2 (dx^2)^2 - a_3^2 (dx^3)^2,
\label{BI}
\end{equation}
with $a_0 = 1$, $x^0 = c t$ and $c = 1$. The metric functions
$a_i$ $(i=1,2,3)$ are the functions of time $t$ only.

Einstein's gravitational field equation for the BI space-time has
the form
\begin{subequations}
\label{BID}
\begin{eqnarray}
\frac{\ddot a_2}{a_2} +\frac{\ddot a_3}{a_3} + \frac{\dot a_2}{a_2}\frac{\dot
a_3}{a_3}&=&  \kappa T_{1}^{1},\label{11}\\
\frac{\ddot a_3}{a_3} +\frac{\ddot a_1}{a_1} + \frac{\dot a_3}{a_3}\frac{\dot
a_1}{a_1}&=&  \kappa T_{2}^{2},\label{22}\\
\frac{\ddot a_1}{a_1} +\frac{\ddot a_2}{a_2} + \frac{\dot a_1}{a_1}\frac{\dot
a_2}{a_2}&=&  \kappa T_{3}^{3},\label{33}\\
\frac{\dot a_1}{a_1}\frac{\dot a_2}{a_2} +\frac{\dot
a_2}{a_2}\frac{\dot a_3}{a_3}+\frac{\dot a_3}{a_3}\frac{\dot
a_1}{a_1}&=&  \kappa T_{0}^{0}. \label{00}
\end{eqnarray}
\end{subequations}
Here $\kappa$ is the Einstein gravitational constant and over-dot
means differentiation with respect to $t$. The energy momentum tensor
for a system of cosmic string and magnetic field in a comoving
coordinate is given by
\begin{equation}
T_{\mu\,(m)}^{\nu} =  \rho u_\mu u^\nu - \lambda x_\mu x^\nu
+ E_\mu^\nu, \label{imperfl}
\end{equation}
where $\rho$ is the rest energy density of strings with massive
particles attached to them and can be expressed as $\rho = \rho_{p} +
\lambda$, where $\rho_{p}$ is the rest energy density of the
particles attached to the strings and $\lambda$ is the tension
density of the system of strings \cite{letelier,pradhan,tade},
which may be positive or negative. Here $u_i$ is the four
velocity and $x_i$ is the direction of the string, obeying the
relation
\begin{equation}
u_iu^i = -x_ix^i = 1, \quad u_i x^i = 0. \label{velocity}
\end{equation}

In \eqref{imperfl} $E_{\mu\nu}$ is the electromagnetic field given by
Lichnerowich \cite{lich}
\begin{equation}
E_\mu^\nu = {\bar \mu} \Bigl[ |h|^2 \Bigl(u_\mu u^\nu - \frac{1}{2}
\delta_\mu^\nu\Bigr) - h_\mu h^\nu \Bigr].
\label{lichn}
\end{equation}
Here $\bar \mu$ is a constant characteristic of the medium and called the magnetic permeability. Typically $\bar \mu$ differs from unity only by a few parts in $10^5$ ($\bar \mu > 1$ for paramagnetic substances and $\bar \mu < 1$
for diamagnetic). In \eqref{lichn} $h_\mu$ is the magnetic flux vector defined by
\begin{equation}
h_\mu = \frac{1}{\bar \mu}  \ast  F_{\nu \mu} u^\nu,
\label{magflux}
\end{equation}
where $\ast F_{\mu\nu}$ is the dual electromagnetic field tensor defined
as
\begin{equation}
\ast F_{\mu \nu} = \frac{\sqrt{-g}}{2} \epsilon_{\mu \nu \alpha \beta}
F^{\alpha \beta}.
\label{dualt}
\end{equation}
Here $F^{\alpha \beta}$ is the electromagnetic field tensor and
$\epsilon_{\mu \nu \alpha \beta}$ is the totally anti-symmetric
Levi-Civita tensor with $\epsilon_{0123} = +1$. Here the co-moving
coordinates are taken to be $u^0 = 1,\, u^1 = u^2 = u^3 = 0$. We
choose the incident magnetic field to be in the direction of
$x$-axis so that the magnetic flux vector has only one nontrivial
component, namely $h_1 \ne 0.$ In view of the aforementioned
assumption from \eqref{magflux} one obtains $F_{12} = F_{13} = 0.$
We also assume that the conductivity of the fluid is infinite. This
leads to $F_{01} = F_{02} = F_{03} = 0$. Thus We have only one
non-vanishing component of $F_{\mu \nu}$ which is $F_{23}.$ Then
from the first set of Maxwell equation
\begin{equation}
F_{\mu\nu;\beta} + F_{\nu\beta;\mu} + F_{\beta \mu; \nu} = 0,
\label{maxe}
\end{equation}
where the semicolon stands for covariant derivative, one finds
\begin{equation}
F_{23} = {\cal I}, \quad {\cal I} = {\rm const.}
\label{f23}
\end{equation}
Then from \eqref{magflux} in account of \eqref{dualt} one finds
\begin{equation}
h_1 = \frac{a_1 {\cal I}}{{\bar \mu} a_2 a_3}.
\label{h1}
\end{equation}
Finally, for $E_\mu^\nu$ one finds the following non-trivial
components
\begin{equation}
E_0^0 = E_1^1 = - E_2^2 = - E_3^3 = \frac{{\cal I}^2}
{2 {\bar \mu} a_2^2 a_3^2}.
\label{E}
\end{equation}

Taking the string along $x^1$ direction and using co-moving
coordinates we have the following components of energy momentum
tensor \cite{ass}:
\begin{subequations}
\label{total}
\begin{eqnarray}
T_{0}^{0} &=&  \rho +
\frac{{\cal I}^2}{2 {\bar \mu}}\frac{a_1^2}{\tau^2}, \label{t00}\\
T_{1}^{1} &=&  \lambda
+\frac{{\cal I}^2}{2 {\bar \mu}}\frac{a_1^2}{\tau^2}, \label{t11}\\
T_{2}^{2} &=&
-\frac{{\cal I}^2}{2 {\bar \mu}}\frac{a_1^2}{\tau^2}, \label{t22}\\
T_{3}^{3} &=&  -\frac{{\cal I}^2}{2 {\bar \mu}}\frac{a_1^2}{\tau^2},
\label{t33}
\end{eqnarray}
\end{subequations}
where we used the definition
\begin{equation}
\tau = a_1 a_2 a_3. \label{taudef}
\end{equation}
It is indeed the volume scale of the BI space-time, i.e.,
$\tau = \sqrt{-g}$ \cite{sahaprd}.

In view of $T_{2}^{2} = T_{3}^{3}$ from \eqref{22}, \eqref{33} one
finds
\begin{equation}
a_2 = a_3 D \ {\rm exp}\Bigl(X \int \frac{dt}{\tau}\Bigr),
\label{b/c}
\end{equation}
with $D$ and $X$ being integration constants. Due to anisotropy of
the source filed, in order to solve the remaining Einstein equation
we have to impose some additional conditions. Here we give two
different conditions. It can be shown that the metric functions can be
expressed in terms of $\tau$. So let us first derive the equation for
$\tau$. Summation of Einstein Eqs. \eqref{11}, \eqref{22}, \eqref{33}
and 3 times \eqref{00} gives
\begin{equation}
\frac{\ddot \tau}{\tau}= \frac{3}{2}\kappa \Bigl(\rho + \lambda +
 \frac{2 {\cal I}^2}{3 {\bar \mu}}\frac{a_1^2}{\tau^2} \Bigr).
\label{dtau1}
\end{equation}
Let us demand the energy-momentum to be conserved, i.e.,
$T_{\mu;\nu}^{\nu} = 0$, which in our case takes the form
\begin{equation}
\frac{1}{\tau}\frac{d}{dt} \bigl(\tau T_0^0\bigr) - \frac{\dot
a_1}{a_1} T_1^1 -\frac{\dot a_2}{a_2} T_2^2  - \frac{\dot a_3}{a_3}
T_3^3 = 0. \label{emcon}
\end{equation}
After a little manipulation from \eqref{emcon} one obtains
\begin{equation}
\dot \rho + \frac{\dot \tau}{\tau}\rho - \frac{\dot a_1}{a_1}\lambda
= 0.
 \label{vep}
\end{equation}
%%%%%%%%%%%%%%%%%%%%%%%%%%%%%%%%%%%%%%%%%%%%%%%%%%%%%%%%%%%%%
\section{Some examples and explicit solutions}
%%%%%%%%%%%%%%%%%%%%%%%%%%%%%%%%%%%%%%%%%%%%%%%%%%%%%%%%%%%%%
In the literature there exists a number of relations between $\rho$ and
$\lambda$, the simplest one being a proportionality relation:
\begin{equation}\label{rhoalphalambda}
\rho = \alpha \lambda
\end{equation}
with the most usual choices of the constant $\alpha$
\begin{equation}\label{alpha}
\alpha =\left \{
\begin{array}{ll}
1 & \quad {\rm geometric\,\,\,string}\\
1 + \omega  & \quad \omega \ge 0, \quad p \,\,{\rm string\,\,\,or\,\,\,
Takabayasi\,\,\,string}\\
-1  & \quad {\rm Reddy\,\,\,string}\,.
\end{array}
\right.
\end{equation}

As it was mentioned earlier, to solve the Einstein equations
completely, we need to impose some additional conditions. Here we
consider the following two.

\subsection{Case 1}
This condition was first used by Bali \cite{bali}.
Following him let us assume that the expansion scalar ($\theta$) in the
model is proportional to the eigenvalue $\sigma_1^1$ of the shear
tensor $\sigma_\mu^\nu$. For the BI space-time we have
\begin{eqnarray}
\theta &=& \frac{\dot a_1}{a_1}+\frac{\dot a_2}{a_2}+\frac{\dot a_3}{a_3}, \label{theta} \\
\sigma_1^1 &=& - \frac{1}{3}\Bigl(4\frac{\dot a_1}{a_1}+\frac{\dot a_2}{a_2}+
\frac{\dot a_3}{a_3}\Bigr). \label{s11}
\end{eqnarray}
Writing the aforementioned condition as
\begin{equation}
\theta = 3 n \sigma_1^1, \label{conbali}
\end{equation}
one comes to the following relation
\begin{equation}
a_1 = Z \bigl(a_2 a_3 \bigr)^N,
\label{a=bc}
\end{equation}
where $N = - (n+1)/(4n+1)$ being the proportionality constant and $Z$
is the integration constant.

 From \eqref{b/c} and \eqref{a=bc} after some
manipulation for the metric functions one finds \cite{ass}
\begin{subequations}
\label{abc}
\begin{eqnarray}
a_1 &=& Z^{1/(N+1)} \tau^{N/(N+1)}, \label{a1}\\
a_2 &=& \sqrt{D}\, \bigl(\frac{\tau}{Z}\bigr)^{1/2(N+1)}\,
{\rm exp} \Bigl[\frac{X}{2} \int \frac{dt}{\tau}\Bigr], \label{a2}\\
a_3 &=& \frac{1}{\sqrt{D}}\,\bigl(\frac{\tau}{Z}\bigr)^{1/2(N+1)}\,
{\rm exp} \Bigl[-\frac{X}{2} \int \frac{dt}{\tau}\Bigr]. \label{a3}
\end{eqnarray}
\end{subequations}
In this case Eq. \eqref{vep} takes the form
\begin{equation}\label{vepcase1}
\dot \rho + \bigl(\rho - \frac{N}{N+1}\lambda \bigr)\frac{\dot
\tau}{\tau} = 0.
\end{equation}
Eq. \eqref{dtau1} now reads
\begin{equation}
\ddot \tau= \frac{3}{2}\kappa (\rho + \lambda) \tau + {\cal X}
\tau^{(N-1)/(N+1)}, \quad {\rm where} \quad {\cal X} = \kappa
\frac{{\cal I}^2}{{\bar \mu}}Z^{2/(N+1)}. \label{dtaucase1}
\end{equation}
We  will see later, the right hand side of \eqref{dtaucase1} is the
function of $\tau$, hence can be written as
\begin{equation}
\ddot \tau = {\cal F}(\tau). \label{tauF}
\end{equation}
Eq. \eqref{tauF} admits first integral which can be written as
\begin{equation}
\dot \tau = \sqrt{2 [{\cal E} - {\cal U}(\tau)]} \label{1stint}
\end{equation}
with
\begin{equation}
{\cal U} (\tau) = - \int {\cal F}(\tau) d\tau. \label{poten}
\end{equation}
The expression \eqref{poten} can be viewed as potential, while ${\cal E}$
as energy level. A detailed analysis of this mechanism can be seen in, e.g.,
\cite{bited}.

Let us now study Eqs. \eqref{vepcase1} and \eqref{dtaucase1} for
different equations of state. Assuming the relation (\ref{rhoalphalambda})
between the pressure of the perfect fluid $\rho$ and the tension
density $\lambda$  from Eq. \eqref{vepcase1} one finds
\begin{equation}
\frac{\dot \rho}{\rho} = (\frac{N}{\alpha (N+1)} -1) \frac{\dot
\tau}{\tau}
\end{equation}
with the solution
\begin{equation}
\rho = C_0 \tau^{\frac{N}{\alpha (N+1)} -1}\,, \label{rhocase1}
\end{equation}
while the equation for $\tau$ reads
\begin{equation}
\ddot \tau= \frac{3}{2} \kappa C_0 (1 + \frac{1}{\alpha})
\tau^{\frac{N}{\alpha (N+1)}} + {\cal X} \tau^{\frac{N-1}{N+1}}.
\label{ddtaucase1}
\end{equation}

This equation can be set in the following form

\begin{equation}
\dot \tau = \sqrt{\frac{3\kappa C_0 (\alpha + 1)(N+1)}{N+\alpha(N+1)}
\tau^{1+N/\alpha(N+1)} + \frac{{\cal X}(N+1)}{N} \tau^{2N/(N+1)} +
{\cal E}_0} \label{vel}
\end{equation}
where ${\cal E}_0$ is the integration constant and is related to
${\cal E}$ as ${\cal E}_0 = 2 {\cal E}$.

\myfiguress{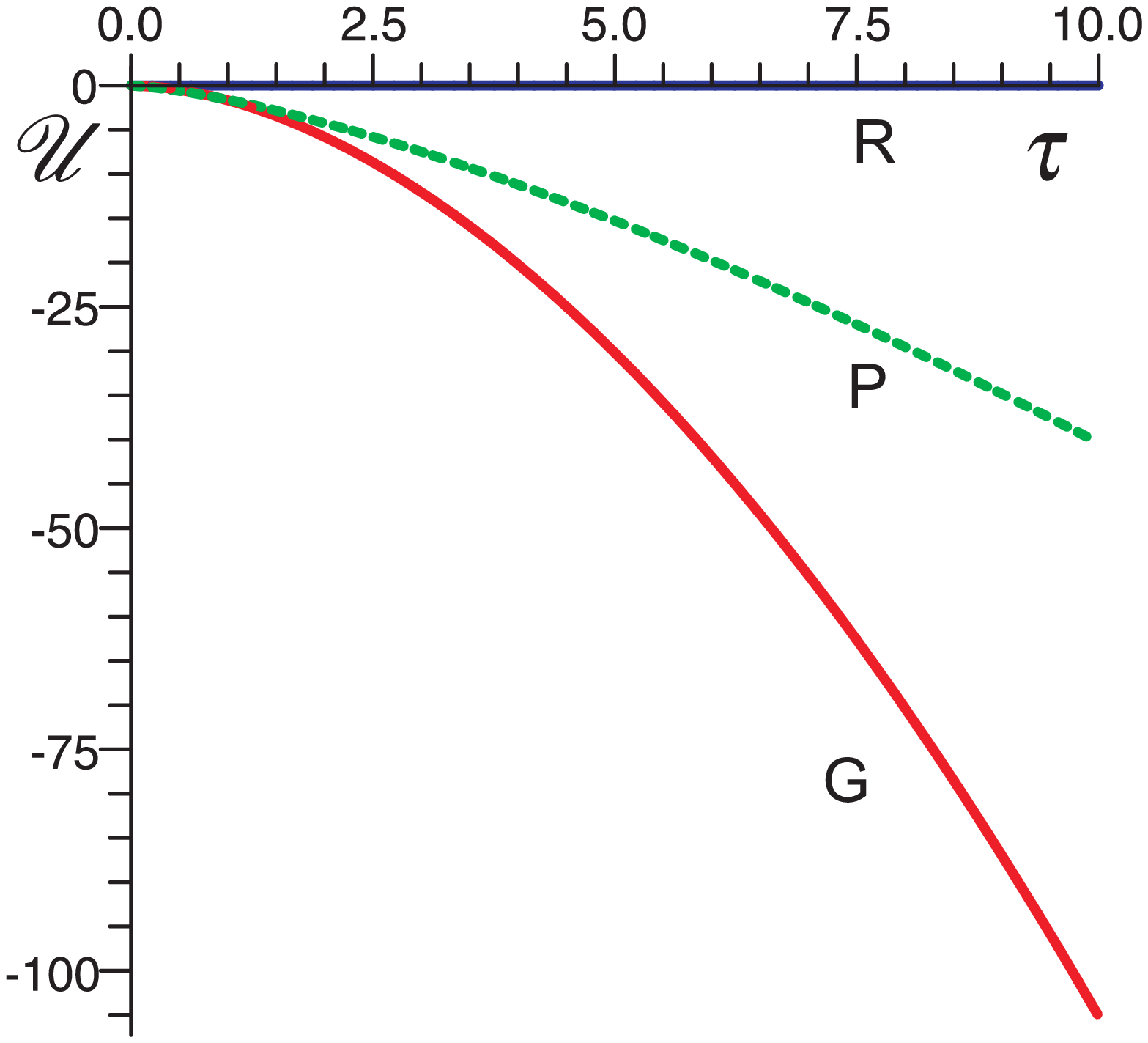}{0.35}{Potential corresponding to the different
equations of state in absence of a magnetic field.}{0.35}{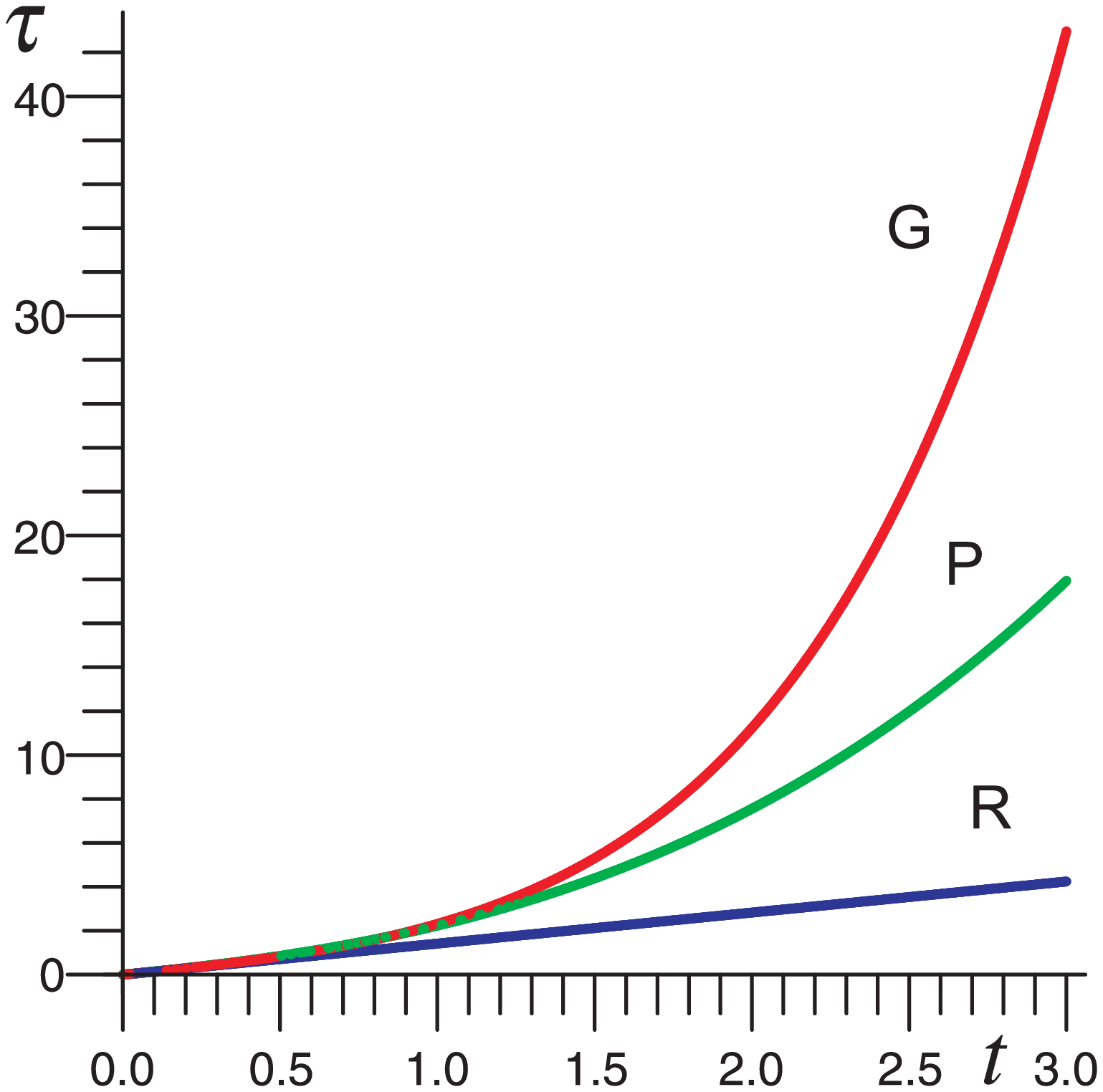}
{0.35}{Evolution of the Universe for different equations of state in
absence of a magnetic field.}{0.35}

\myfiguress{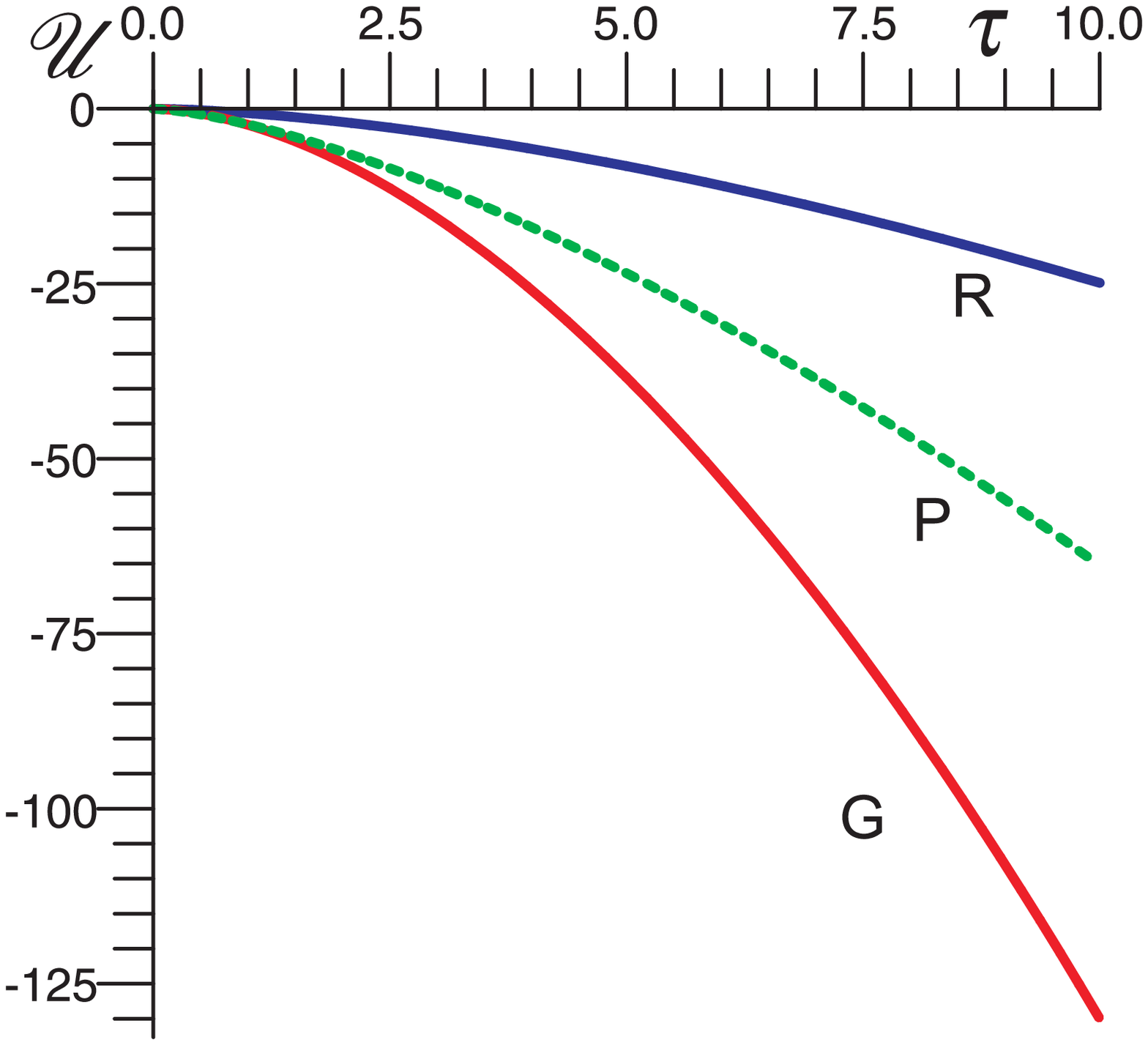}{0.35}{Potential corresponding to the different
equations of state in presence of a magnetic field.}{0.35}
{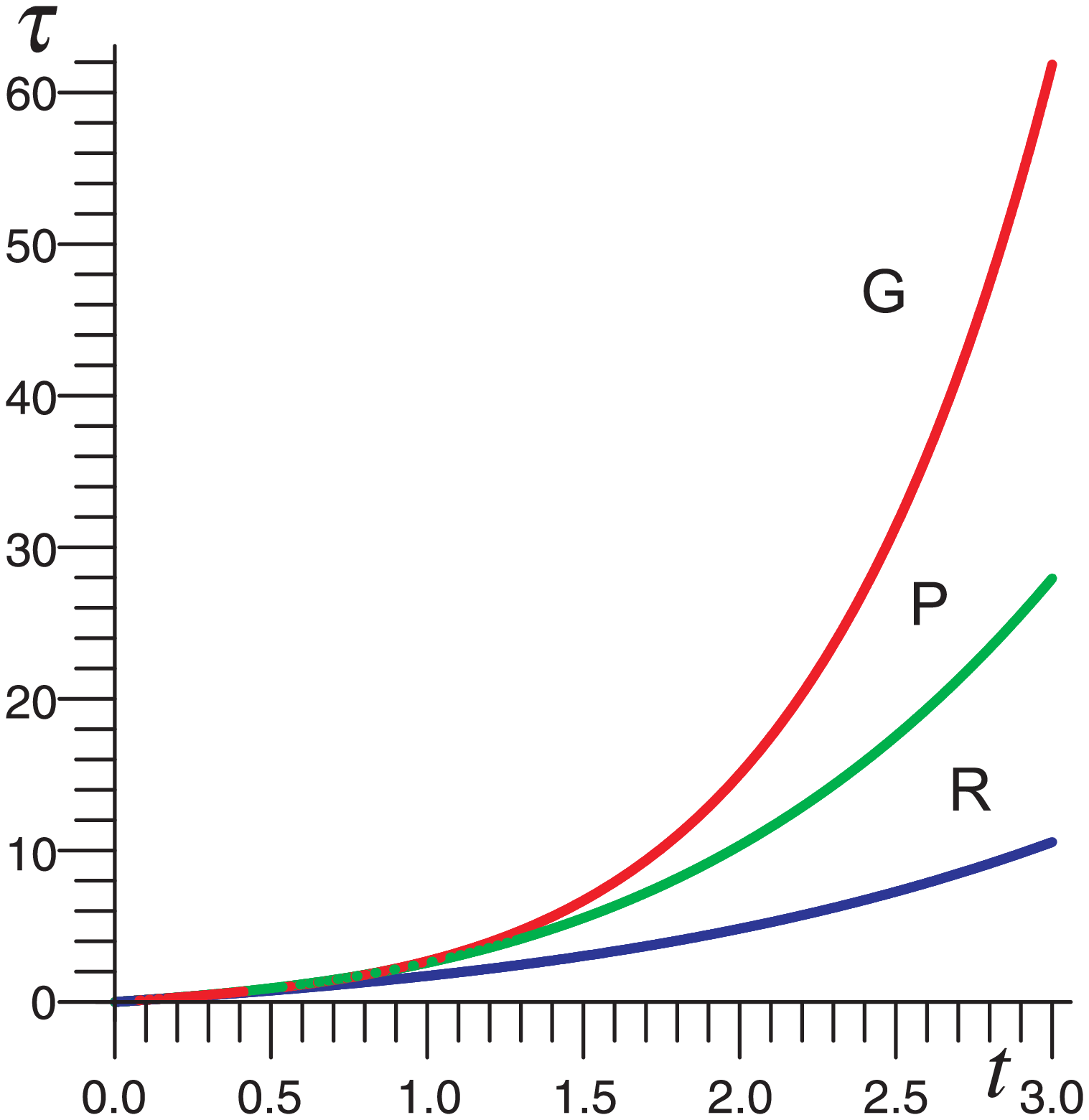}{0.35}{Evolution of the Universe for different equations of
state in presence of a magnetic field.}{0.35}

In Figs. \ref{potencs30} and \ref{potencs3} we have illustrated the
potential corresponding to the different equations of state without
and with the magnetic field, respectively. The Figs. \ref{taucs30}
and \ref{taucs3} show the evolution of $\tau$ for different cases.
Here "G", "P" and "R" stand for geometric string, $p$ string and
Reddy string, respectively. The reason to illustrate figures with and
without magnetic field is to show the role of magnetic field. As one
sees, in all cases $\tau$ might be zero at the initial stage of
evolution, thus giving rise to the initial singularity in one hand,
$\tau$ is not bound from above which means in all three cases we have
ever expanding Universe. But introduction of magnetic field into the
system results in rapid growth of $\tau$. In numerical analysis we
used the following value for the problem parameters: $\kappa = 1$,
$\omega = 1$, $N=4$, $Z = 1$.  In case of string only we set ${\cal I} = 0$, otherwise ${\cal I} = 1$. For magnetic permeability we choose $\bar \mu = 1.00001$ and $\bar \mu = 0.99999$, respectively. Since it doesn't make any significant change in the behavior of $\tau$, we illustrate only case with $\bar \mu = 1.00001$. The initial value for $\tau$ is taken to be $\tau (0) = 0.0001$ and corresponding first derivative is calculated from \eqref{vel} at ${\cal E}_0 = 1$.

\subsection{Case 2}
The second case was proposed by Barrow \cite{barrow}.
Here we first introduce the generalized Hubble parameter
\begin{equation}
H = \frac{1}{3}\bigl(\frac{\dot a_1}{a_1}+ \frac{\dot a_2}{a_2} +
\frac{\dot a_3}{a_3} \bigr) = \frac{1}{3}\frac{\dot \tau}{\tau}, \label{Hubble}
\end{equation}
and two relative shear anisotropy parameters by
\begin{subequations}
\begin{eqnarray}
R &=& \frac{1}{H}\bigl(\frac{\dot a_1}{a_1} - \frac{\dot a_2}{a_2}\bigr), \label{sp1}\\
S &=& \frac{1}{H}\bigl(\frac{\dot a_1}{a_1} - \frac{\dot a_3}{a_3}\bigr).
\label{sp2}
\end{eqnarray}
\end{subequations}
When $R = S = 0$ the universe will be the isotropic flat Friedman universe.
If one sets $R = {\rm const.}$, then in view of \eqref{Hubble} one finds,
\begin{equation}
a_1 = D_0 a_2 \tau^{R/3}. \label{a/b}
\end{equation}
Then, in view of \eqref{taudef}, \eqref{b/c} we find the following expressions for metric functions
\begin{subequations}
\label{abcb}
\begin{eqnarray}
a_1 &=& (D_0^2 D)^{1/3} \tau^{(3+2R)/9} e^{\frac{X}{3}\int\frac{dt}{\tau}}, \label{a1b}\\
a_2 &=& (D_0^{-1} D)^{1/3} \tau^{(3-R)/9} e^{\frac{X}{3}\int\frac{dt}{\tau}}, \label{a2b}\\
a_3 &=& (D_0 D^2)^{-1/3} \tau^{(3-R)/9} e^{\frac{-2X}{3}\int\frac{dt}{\tau}}. \label{a3b}
\end{eqnarray}
\end{subequations}
In this case Eq. \eqref{vep} takes the form
\begin{equation}
\dot \rho + \bigl(\rho - \frac{3+2R}{9}\lambda \bigr)\frac{\dot
\tau}{\tau} - \frac{X}{3\tau}\lambda = 0\,,
 \label{vepcase2}
\end{equation}
whereas, for $\tau$ in this case we have
\begin{equation}
\frac{\ddot \tau}{\tau}= \frac{3}{2}\kappa \Bigl(\rho + \lambda +
 \frac{2 {\cal I}^2}{3 {\bar \mu}}(D_0^2D)^{2/3}
\tau^{(2R-15)/9}e^{\frac{2X}{3}\int\frac{dt}{\tau}} \Bigr).
\label{dtau1barrow}
\end{equation}

Comparing this equation with the corresponding one (\ref{ddtaucase1}) from
{\bf Case 1}, it results that in {\bf Case 2} the equation of evolution
for $\tau$ is more intricate. Indeed Eq. \eqref{dtau1barrow} involves
the function $\tau$ in an integral and needs further study. A detailed
analysis is presently under investigation and will be reported in a
forthcoming paper \cite{SV}.

In the present paper we confine ourselves to a preliminary task
considering this equation in an asymptotic regime, for
large $t$. To be aware of possible asymptotic behavior of $\rho$ and
$\tau$ for large $t$, let us assume again relation (\ref{rhoalphalambda})
and Eq. (\ref{vepcase2}) becomes
\begin{equation}\label{rtcase2}
\frac{\dot \rho}{\rho} = (\frac{3 + 2 R}{9\alpha} - 1) \frac{\dot
\tau}{\tau} + \frac{X}{3 \alpha}\frac{1}{\tau} \,,
\end{equation}
which shows that both sides of this equation are the same function of
time. As in the {\bf Case 1} Eq. (\ref{rtcase2}) allows a behavior
of the pressure of the perfect fluid $\rho$  decreasing in time as an
inverse power of $t$:
\begin{equation}\label{rhocase2}
\rho = C_1 t^{-\beta}\,,
\end{equation}
$C_1$ and $\beta\geq 0$ being some constants.
Correspondingly, for $\tau$ we get  the equation
\begin{equation}\label{taucase2}
(\frac{3+2R}{9\alpha} - 1) \dot\tau + \frac{\beta}{t}\tau +
\frac{X}{3 \alpha} =0
\end{equation}
with the solution
\begin{equation}\label{tau2sol}
\tau = - \frac{3X}{9 \alpha \beta + 3 + 2 R - 9\alpha} t +
C_2 t^{-\frac{9\alpha \beta}{3 + 2R -9\alpha}}\,.
\end{equation}

It is interesting to observe that Eq. (\ref{rtcase2}) allows even an
exponential behavior of $\rho$ for large $t$:
\begin{equation}\label{rhoexp}
\rho = C_3 e^{-\gamma t}
\end{equation}
which means that the both sides of Eq. (\ref{rtcase2}) are a negative
constant $(-\gamma)$. The corresponding behavior of $\tau$ from
Eq. (\ref{rtcase2}) is
\begin{equation}\label{tauexp}
\tau = - \frac{X}{3\alpha\gamma} + C_4 e^{-\frac{9\alpha \gamma }
{3 + 2 R - 9 \alpha}t}
\end{equation}
with $C_4$ another arbitrary constant.

The full study of Eq. (\ref{dtau1barrow}), which contains the parameters
characterizing the cosmic string and magnetic field, for large $t$ using
the asymptotic behaviors (\ref{rhocase2}), (\ref{taucase2}) and
(\ref{rhoexp}), (\ref{tauexp}) will be presented elsewhere \cite{SV}.
%%%%%%%%%%%%%%%%%%%%%%%%%%%%%%%%%%%%%%%%%%%%%%%%%%%%%%%%%%%%%%%%%%%%%%%%
            \section{Conclusions}
%%%%%%%%%%%%%%%%%%%%%%%%%%%%%%%%%%%%%%%%%%%%%%%%%%%%%%%%%%%%%%%%%%%%%%%%
In the present paper we investigated in the frame of Bianchi type I
models a string cosmological model in the presence of a magnetic field.
We used some tractable assumptions concerning the parameters entering
the model.

In the case of a proportionality between the trace of the expansion
tensor $\theta$ and the eigenvalue $\sigma_1^1$ of the shear tensor we
are able to get explicit analytic solutions. Setting the relative
anisotropy parameter $R= const$ the model is more involved, but
sufficiently interesting to deserve further study.

\begin{acknowledgments}
This work is supported in part by a joint Romanian-LIT, JINR, Dubna Research
Project, theme no. 05-6-1060-2005/2010.
\end{acknowledgments}

%%%%%%%%%%%%%%%%%%%%%%%%%%%%%%%%%%%%%%%%%%%%%%%%%%%%%%%%%%%%%%%%%%%%%%%%%%%

\newcommand{\hnl}{\htmladdnormallink}


\begin{thebibliography}{99}
%1
\bibitem{KM} Koivisto T. and Mota D.F., Phys. Lett. B, {\bf 644},
104 (2007); Koivisto T. and Mota D.F. , Phys. Rev. D, {\bf 75},
023518 (2007).
%2
\bibitem{GR} Grasso D. and Rubinstein H.R., Phys. Rep., {\bf 348},
163 (2001).
%3
\bibitem{W} Widrow L.M., Rev. Mod. Phys., {\bf 74}, 775 (2002).
%4
\bibitem{letelier} Letelier, P.S., Phys. Rev. D, {\bf 28}, 1424
(1983).
%5
\bibitem{pradhan} Pradhan A., Yadav A.K., Singh R.P. and Singh V.K.,
Astrophys. Space Sci., {\bf 312}, 145 (2007).
%6
\bibitem{tade} Khadekar G.S., Tade S.D., Astrophys. Space Sci. {\bf 310}, 47
(2007)].
%7
\bibitem{lich} Lichnerowicz A.,
{\it Relativistic Hydrodynamics and Magnetohydrodynamics},
(Benjamin, New York, 1967).
%8
\bibitem{ass} Saha Bijan, J. Astrophys. Space Sci. 2005. {\bf 299}
No. 1, \hnl{149-158}{http://www.jinr.ru/~bijan/my_papers/ass_149.pdf},
(2005).
%9
\bibitem{sahaprd} Saha Bijan, Phys. Rev. D. {\bf 64}
\hnl{123501.}{http://www.jinr.ru/~bijan/my_papers/PRD23501.pdf} (2001).
%10
\bibitem{bali} Bali R. Int. J. Theor. Phys.
{\bf 25}, 755 (1986).
%11
\bibitem{bited} Bijan Saha and Todor Boyadjiev, Phys. Rev. D {\bf 69},
\hnl{124010}{http://www.jinr.ru/~bijan/my_papers/PRD24010.pdf}, (2004)
[arXiv:\,\hnl{gr-qc/0311045}{http://xxx.lanl.gov/abs/gr-qc/0311045}].
%12
\bibitem{barrow} Barrow J.D., Phys. Rev. D. {\bf 55} 7451 (1997).
%13
\bibitem{SV} Saha B. and Visinescu M., in preparation.

\end{thebibliography}
\end{document}